# Scalable manufacturing of polarization-insensitive metalenses with high-uniform focal arrays in the visible

*Xu Mao[1,3,4,†,*], Gang Yu[2,†], Hongsheng Ding[2,†], Yongmei Zhao[1,3], Chaowei Si[1], Fuhua Yang[1,3,4], and Xiaodong Wang[1,3,4]*

[1]*Engineering Research Center for Semiconductor Integrated Technology, Institute of Semiconductors, Chinese Academy of Sciences, Beijing 100083, China*
[2]*Department of Physics, School of Mathematics and Physics, University of Science and Technology Beijing, Beijing 100083, China*
[3]*School of Integrated Circuits & Center of Materials Science and Optoelectronics Engineering, Chinese Academy of Sciences, Beijing 100049, China*
[4]*Beijing Engineering Research Center of Semiconductor Micro-Nano Integrated Technology, Beijing 100083, China*

[†]These authors contributed equally to this work.
[*]E-mail: maoxu@semi.ac.cn

**Abstract**

Multi-foci metalenses with uniform focal arrays attract special attention as they enable a single incident beam to focus on the same focal plane and share the identical numerical apertures. In this work, we demonstrate the scalable manufacturing of polarization-insensitive metalenses with high-uniform focal arrays in the visible. To overcome the limitations inherent in conventional imprint resins characterized by a low refractive index, hybrid meta-atoms are formed using the imprint resin itself, and a thin layer of titanium dioxide ($TiO_2$) is deposited on it via atomic layer deposition (ALD) to increase the effective refractive index, enabling full phase coverage. We propose the gradient-descent-optimization strategy for phase retrieval of multi-foci metalens, which renders our design free from cross-talk and makes it feasible to achieve uniform focal arrays with flexible geometries. Based on this inverse-design scheme, the capability of nanoimprinted metalenses are explored by directly producing various focal arrays, including complex geometries such as square, rhombic lattices, as well as intact and defective rings. We envision this work may pave the way for various fields, including optical trapping, materials science, and quantum optics.

# 1. Introduction

Metasurfaces, composed of subwavelength arrays of artificially designed structures (known as meta-atoms), have generated significant research interest in optical research. They offering several advantages over bulk optics, including compact size, subwavelength spatial resolution, and flexible polarization control[1-4]. Benefiting from the unprecedented capability of metasurfaces in light control, metalenses are ultrathin and lightweight, and can be designed to versatile functionalities. The precise manipulation of the focal spots (e.g., size, focal length, and number) is a critical research topic in the field of metalenses. High-numerical-aperture metalenses with close to diffraction-limited focal spots have been utilized for subwavelength imaging[5], stable trapping[6], and coupling of free-space beams to photonic devices[7]. The achromatic[8,9] and wide-angle[10] metalenses, respectively maintaining the same focal length across various wavelengths and incident angles, showcase impressive working bandwidth and field of view. Among these advancements, multi-foci metalenses with uniform focal arrays attract special attention as they enable a single incident light beam to focus on the same focal plane and share the identical numerical apertures[11,12]. Utilizing this unique functionality, a miniaturized and simplified two-photon polymerization (TPP) system was developed and demonstrated efficient multi-focus parallel processing[13]. Wavelength and polarization information can be independently encoded into each focal point, enabling the development of compact spectrometer[14] and polarization-resolved device[15]. More importantly, metalens-based optical trapping arrays have emerged as a powerful platform for the preparation and manipulation of ultracold atoms[16,17], enabling multifunctional control in complex quantum information experiments.

One popular scheme to achieve multi-foci functionality is based on physical intuition to incorporate parts of individual single-foci metalenses concentrically into a large metalens[18,19], which has significant challenges to achieve the desired multi-foci effect due to the substantial cross-talk among the unit cells, as well as the consequent sacrifice of the effective aperture of each sub-metalens. A similar intuition-guided design method, using multiple identical metalenses to form a metalens array, can create multiple focal points in a 2D plane, providing the same focusing performance[20,21]. However, the numerical apertures (NAs) are limited, and the spacing between focal points is discrete. Alternative design strategy have been proposed by consolidating the phase profiles of individual metalenses into a single phase profile[22](phase superposition method), which works well only when the number of focal points is constrained. Although iterative phase retrieval algorithms, such as the Gerchberg–Saxton (GS) algorithm[23], have been utilized to achieve uniform focal point arrays, the design of efficient metalenses with high positioning accuracy, as well as size and intensity uniformity of the generated focusing traps still remains a challenge.

On the other hand, fabricating large-area metasurfaces, especially for the visible or shorter-wavelengths metasurfaces, presents a significant challenge for massive productions. The main patterning technique is limited to high-resolution electron-beam lithography (EBL)[24], which is expensive, time-consuming, and limited patterning areas. Although photolithography is capable of overcoming these limitations in patterning due to its capability to fabricate large-scale and mass-scalable metasurfaces[25], several challenges remain, such as inadequate patterning resolution, material constraints, and high manufacturing costs. Fortunately, nanoimprint lithography (NIL), as a highly advanced and cost-effective manufacturing technique for nanostructures, has proven to be a promising alternative to fabricate meta-device in the visible[26,27]. The current nanoimprint lithography works have primarily concentrated on Pancharatnam-Berry phase metasurfaces with uniform unit cell dimensions[28], which are feasible to fully fill resin and complete release during demold process. However, these types of metasurfaces solely perform effectively under circularly polarized light, greatly limiting for various practical applications. Our work specifically emphasizes the scalable manufacturing of polarization-dependent metasurfaces composed of high-aspect-ratio nanopillars with varying radius. The nanoimprint lithography multi-foci metalenses benefits from the combined advantages of metasurface and nanoimprint technology, enabling high-throughput meta-devices at a low cost manner and offering possibilities of precise focal spot control.

In this work, we demonstrate the scalable manufacturing of polarization-insensitive metalenses with high-uniform focal arrays in the visible, in which hundreds of metalenses can be fabricated at scale once a single master mold is imprinted. We simplify the fabrication process by using the imprint resin itself as the meta-atoms and depositing a thin layer of titanium dioxide ($TiO_2$) on the it via atomic layer deposition (ALD) to increase the effective refractive index, providing full phase coverage. We propose the gradient-descent-optimization strategy for phase retrieval of multi-foci metalens, which renders our design free from cross-talk and makes it feasible to achieve uniform focal arrays with flexible geometries. Based on this inverse-design scheme, the capability of nanoimprinted metalenses are explored by directly producing various focal arrays, including complex configurations such as square, rhombic lattices, as well as intact and defective rings. Furthermore, multifocal imaging is performed to investigate performance of the multi-foci metalenses. We envision this work may pave the way for various fields, including optical trapping, materials science, and quantum optics.

## 2. Results and Discussion

The high-uniform multi-foci metalens, composed of polarization-insensitive and $TiO_2$-coated hybrid nanopillars, is specially designed to operate at the wavelength of 450 nm, a common wavelength for blue light, offering a prospective way to realize high-resolution microscopy and parallel laser processing (**Figure 1a**). The conformal layer of titanium dioxide ($TiO_2$) is coated onto the imprinted resin nanostructures via atomic layer deposition (ALD), which could increase the effective refractive index ($n_{eff}$) of hybrid nanopillars and effectively achieve sufficient phase coverage. Here, $TiO_2$ is selected as the coating material owing to its high refractive index ($n$) and low extinction coefficient ($k$) in the visible region (**Supplementary Note 1**). To clarify this key point, transverse electric (TE) eigenmodes of the nanopillars without (**Figure 1b**) and with $TiO_2$ coverage (**Figure 1c**) are calculated by finite element method (FEM). High-index $TiO_2$ film surrounding the resin confines the electric field and prevents leakage into free space. The effective refractive index of TE mode for hybrid nanopillar (nanopillar radius: 50 nm, $TiO_2$ thickness: 22 nm) is 1.142, whereas in the absence of $TiO_2$, it is only 1.025. Through this ALD-$TiO_2$ method, the limitation of low refractive index ($n \approx 1.5$) in most existing imprinted resin can be effectively addressed, paving the way for the commercialization of meta-optics. As a proof of concept, we demonstrate the scalable manufacturing of visible polarization-insensitive metalenses with high-uniform focal arrays (**Figure 1d**).

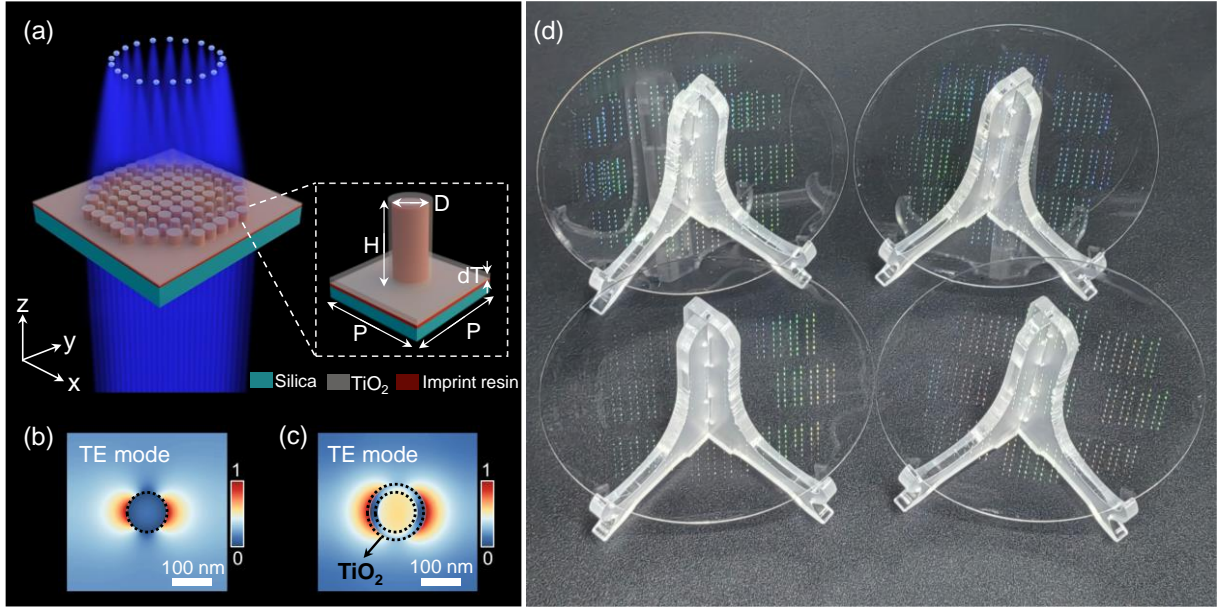

**Figure 1**. (a) Schematic of the high-uniform multi-foci metalens with hybrid nanopillars consisting of the imprint resin and high-index $TiO_2$ thin film. (b, c) Transverse electric (TE) eigenmodes of the nanopillars without (b) and with $TiO_2$ coverage (c). (d) Photograph of mass-produced multi-foci metalenses on 4″ wafers.

To determine the thickness $dT$ of the thin $TiO_2$ film, we employed the finite-difference time domain (FDTD) method to calculate the phase shift and transmission for different configurations of hybrid nanopillars (**Figure 2a**). Here, meta-atoms (made of only imprint resin) with a height of H = 900 nm and a period of P = 400 nm are specially selected, featuring diameters ranging from 80 nm to 320 nm to ensure a minimum linewidth and spacing of 80 nm and aspect ratios of approximately ten, considering the manufacturing feasibility. The phase range of the hybrid meta-atoms increases with the thickness of $TiO_2$ film, but this is accompanied by a reduction in transmission. The thickness $dT$ = 22 nm is ultimately determined, which ensures that eight hybrid meta-atoms achieve complete $0\sim2\pi$ phase coverage with an average transmission close to 70%, whereas without $TiO_2$, the phase coverage is limited to around 0 to $\pi$ (**Figure 2b**).

We implement an inverse design method based on gradient-descent (GD) optimization to retrieve the phase profile of the multi-foci metalens to generate flexible focal point arrays with high positioning accuracy, as well as size and intensity uniformity (**Figure 2c**). Our optimization begins with customized focal arrays composed of diffraction-limited Airy disks, which serve as the target electric field ($|\widehat{U}|^2$). The metalens plane is discretized into M × N unit cells with a period P, each initialized with a uniform amplitude $A_0 = 1$ and a fixed phase $\varphi_0 = \pi$. The reconstructed electric field ($|U_z|^2$) at focal plane can be obtained by angular spectrum method (ASM) as follows[29]:

$$U_z(x, y, z_0) = \mathcal{F}^{-1}\{\mathcal{F}[U(x, y, 0)] \times H(\xi, \eta, z_0)\} \quad (1)$$

where $H$ is the optical transfer function of wave propagation in free space. $\xi$ and $\eta$ are spatial frequencies. $\mathcal{F}^{-1}$ and $\mathcal{F}$ are the inverse Fourier transform and Fourier transform, respectively. Phase $\varphi_0$ is considered as an optimization variable, and the phase is retrieved to minimize the loss function $\mathcal{L}$. This function employs the mean squared error (MSE) to quantitatively evaluate the fidelity between the target and reconstructed focal arrays. Notably, we assign different weights to the focal and non-focal loss terms based on specific features to simultaneously achieve low sidelobes, high uniformity, and high positioning accuracy (**Supplementary Note2**):

$$\mathcal{L} = w_1 * MSE_{focus} + w_2 * MSE_{non-focus} \quad (2)$$

For gradient descent optimization, an adaptive moment estimation (Adam) is implemented as the optimization algorithm (**See Experimental Section for details**). With the assistance of the GD optimization, $\varphi_0$ can be updated and sent to the next iteration. The values of the individual loss terms have reached a plateau after 800 iterations (**Figure 2d**), and the optimized phase mask $\varphi_{ml}$ is output. The multi-foci metalens layout is formed by mapping $\varphi_{ml}$ to the corresponding hybrid meta-atoms. We have compared the performance of our inverse design method with the common phase superposition method for generating focal arrays (**Figure 2e, 2f**). The results demonstrate that the focal arrays reconstructed by our method exhibits a distribution profile consistent with the target Airy disks arrays, with peak intensities of all focal spots approaching 1. In contrast, the focal arrays obtained using the direct phase superposition method (**Supplementary Note3**) exhibit significant sidelobes and peak intensities below 0.3.

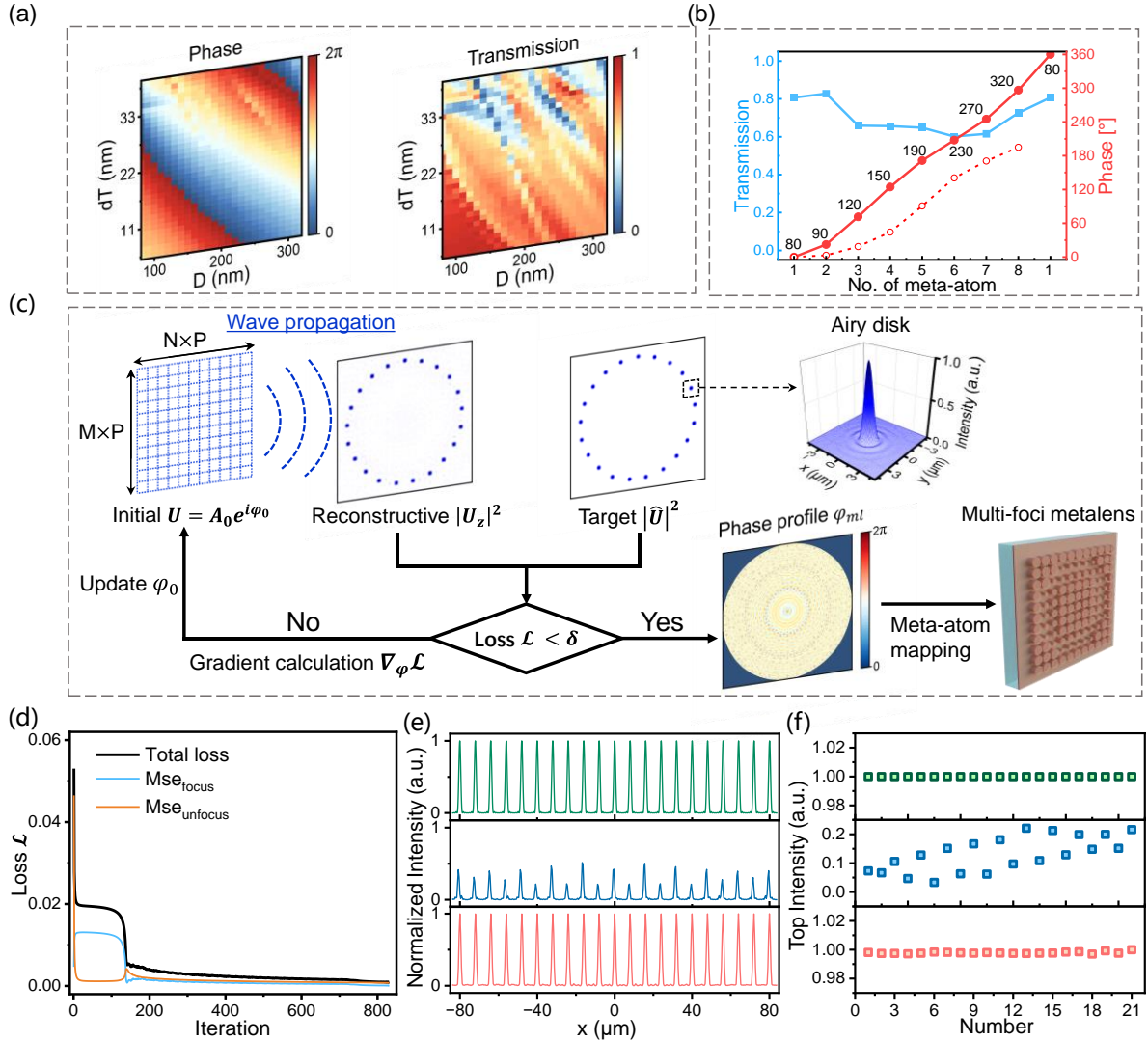

**Figure 2**. (a) The phase shift and transmission for different configurations of hybrid nanopillars. (b) Transmission (blue squares) and phase (red circles) as a function of the index for the eight hybrid meta-atoms with different diameters. The red dashed line represents the phase shift of only the imprint resin. The diameter D values (made of only imprint resin) are marked along the phase distribution line, next to the red circles. (c) Flowchart of the inverse design method for high-uniform multi-foci metalens. (d). Loss value as a function of the iteration number in the gradient descent optimization algorithm. (e) The normalized intensity distributions and (f) the peak intensities along a 10 μm horizontal line centered on each focal spot for the target Airy disks (green), using phase superposition method (blue), and using our method (red).

Our proposed NIL process involves the fabrication of the master and the soft mold (These two are replica molds, which can be recycled and reused), and transferring the metasurface pattern from the soft mold to the substrate by UV-curing nanoimprint (**Figure 3a**). Specifically, we have fabricated nine groups of metasurface arrays across different regions of a 4-inch silica wafer to determine the nanoimprinting uniformity and yield (**Figure 3b, 3c**). The master mold consists of nanopillars with a height of 900 nm, which are patterned by EBL and etched into a silicon substrate by inductively coupled plasma reactive ion etching (**Figure 3d, Supplementary Note4**). During soft mold fabrication, the master mold is first fluorinated to reduce adhesion in order to transfer the nanostructures with high fidelity while avoiding the breaking of high aspect ratio features. After that, the PS05 soft mold resin is spin-coated onto the master mold, and a polyethylene terephthalate (PET) plate is placed on top to support the PS05 layer. They are then placed in the nanoimprint device (Uniprinter), where pressure is applied and they are irradiated by a UV source. Subsequently, the soft mold is released from the master mold and can be used directly for imprinting samples (**Figure 3e**). Notably, to enhance the adhesion between the nanoimprint resin (V50) and substrate, the sol−gel binder (C-primer adhesive) is firstly spin-coated on the silica. After completing these steps, the sample is placed back into the Uniprinter and then imprinted using the soft mold. After being fully cured under pressure and UV-light irradiation, the demolding process can be carried out. Finally, a high-index 22 nm $TiO_2$ film is thinly deposited on the imprinted structures (**Figure 3f**) using ALD to achieve full phase modulation. The optical microscope and scanning electron microscope (SEM) are utilized to characterize the shape and dimensions of the fabricated samples (**Figure 3b-3f**). It can be seen that the nanostructures fabricated by NIL process exhibit good circular symmetry, with no observable missing nanopillars. The high aspect ratio and varying filling factors of the nanopillars exhibit good verticality and minimal deformation.

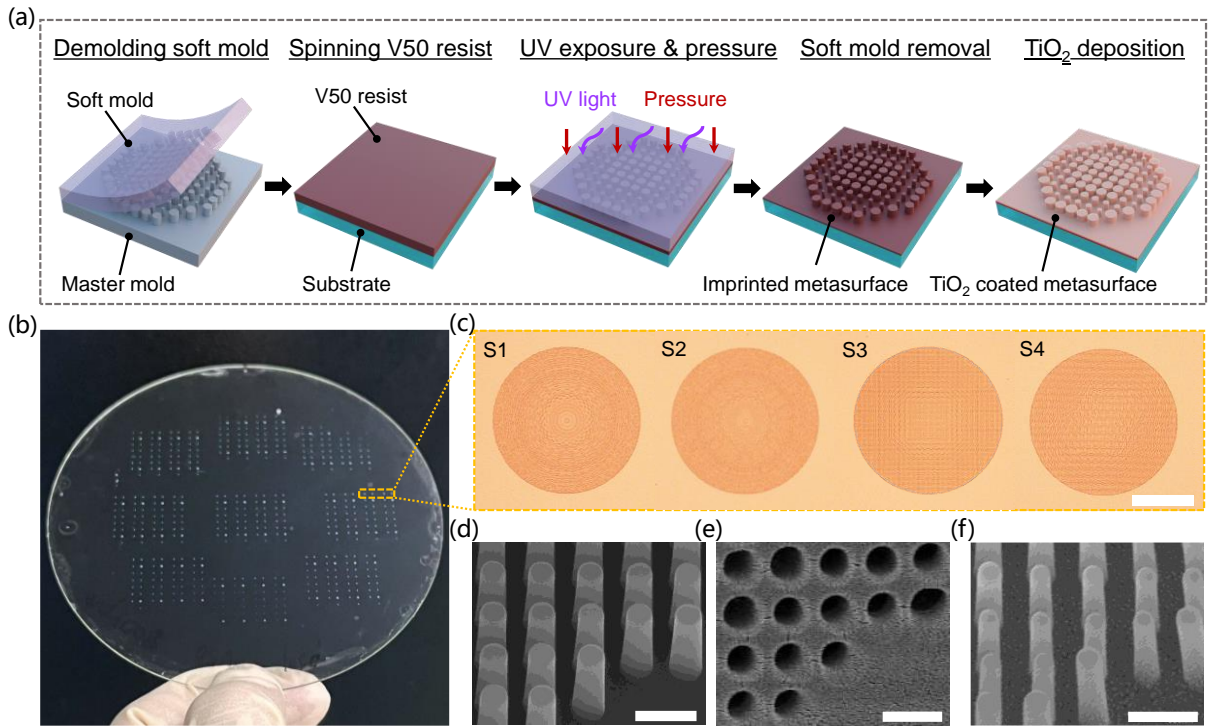

**Figure 3**. (a) Schematic of the massive production of high-index visible polarization-insensitive metasurface. (b) Nine groups of metasurface arrays across different regions of a 4-inch silica wafer. (c-f) Optical microscope and SEM characterizations of the fabricated muti-foci metalenses. (c) The optical microscope images of metalenses with the intact ring (S1), defective ring (S2), square lattice (S3), and rhombic lattice (S4). Scale bar: 100 μm. SEM images of (d) the master mold, (e) the soft mold, and (f) the metalens sample. All scale bars: 500 nm.

The performance of the fabricated muti-foci metalenses is evaluated by the characterization of the focusing properties using a customized measurement apparatus (**Supplementary Note 5**). We have experimentally realized a number of focal arrays to demonstrate the ability of metasurfaces to generate arbitrary configurations. Our demonstrated 2D arrays include an intact ring array with 7.2 μm spacing (**Figure 4a**), a defective ring array with 7.2 μm spacing (**Figure 4b**), a square lattice with a lattice constant of 8.4 μm (**Figure 4c**), and a rhombic lattice with a lattice constant of 8 μm (**Figure 4d**).

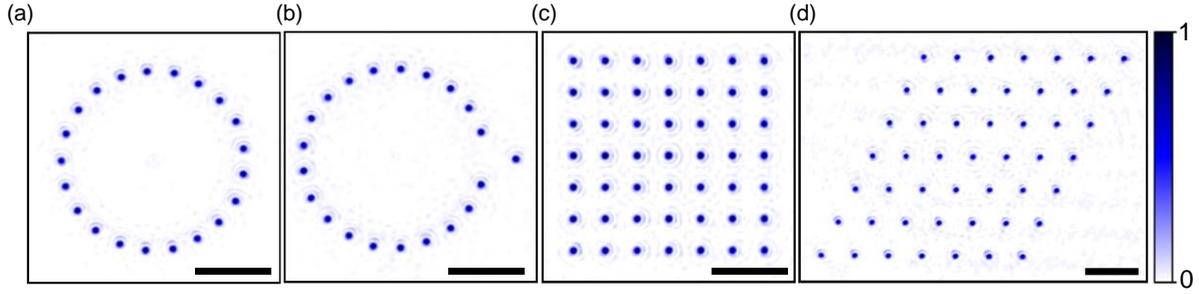

**Figure 4**. Measured intensity profiles of (a) an intact ring array with 7.2 μm spacing, (b) a defective ring array with 7.2 μm spacing, (c) a square lattice with a lattice constant of 8.4 μm, and (d) a rhombic lattice with a lattice constant of 8 μm. The intensity profiles in (a-d) are normalized to their peak value. All scale bars: 20 μm.

## 3. Conclusion

In summary, we demonstrate the scalable fabrication of polarization-insensitive metalenses capable of generating high-uniformity focal arrays in the visible spectrum. We introduce a gradient-descent optimization strategy for phase retrieval in multi-foci metalenses, eliminating cross-talk and enabling the realization of uniform focal arrays with flexible geometries. Leveraging this inverse-design approach, the potential of nanoimprinted metalenses is explored through the direct production of diverse focal arrays, including complex configurations such as square and rhombic lattices, as well as intact and defective ring structures.

**Acknowledgements**

This work was supported by the National Natural Science Foundation of China (No. 62274152)